\documentclass[useAMS,usenatbib]{mn2e} 
\usepackage{times}
\usepackage[dvips]{graphicx} 

\title{Spin-down age: the key to magnetic field decay}
\author[A.P. Igoshev]{A.P.~Igoshev$^1$\thanks{ignotur@gmail.com}\\
$^1$Astrophysical Institute of Saint-Petersburg University, Universitetskyi pr. 28, Staryi Peterhof, Saint-Petersburg, Russia, 198504
}

\date{Released 2012 Xxxxx XX}

\pagerange{\pageref{firstpage}--\pageref{lastpage}} \pubyear{2002}

\def\LaTeX{L\kern-.36em\raise.3ex\hbox{a}\kern-.15em
    T\kern-.1667em\lower.7ex\hbox{E}\kern-.125emX}

\begin{document}

\maketitle
\label{firstpage}

\maketitle

\begin{abstract}
The properties of the spin-down age are investigated. Based on assumption about a uniform magnetic field decay law we suggest a new method which allows us
to shed light on magnetic field decay. This method is applied for following selection:
isolated non-millisecond pulsars from the ATNF catalog are chosen. Pulsars in the selection are with the spin-down ages from $4\cdot 10^4$ to $2\cdot 10^6$ years.
In order to avoid observational selection we take into account only pulsars which are closer to the Sun than 10 kpc. For this selection we restore the uniform magnetic field decay 
law. It appears that the magnetic field decays three times from $4\cdot  10^4$ to $3.5\cdot 10^5$ years. This function is approximated by modified power-law. 
We also estimate the birthrate of pulsars in our Galaxy and find that it should be about 2.9 pulsars per century.
\end{abstract}

\section{Introduction}

The spin-down ages of radio-pulsars are determined by values of $P$ - periods and $\dot P$ - periods derivative.
They change with time. If n is the braking index and n=3,  these characteristics evolve due to $P\dot P=\mathrm{const}$ law. 
However, we observe different braking indexes in the nature. Therefore we decide to investigate a uniform function $f(t) = \sqrt{P\dot P}/\alpha B_0$.
This function is not a constant because of magnetic field decay, alignment or some combination of both these phenomena. 
Although we suppose that $f(t)$ changes because of magnetic field decay, this assumption does not affect generality of our results.

Indeed, there are a few evidences both theoretical and observational which indicate that magnetic field of pulsars should decay. 
The theoretical evidence is that the current keeping the magnetic field should decrease because of following reasons: 
finite conductivity of matter in neutron stars (ohmic dissipation), the Hall effect or ambipolar diffusion
\citep{Goldreich1992}.
The observational evidence is that temperature of neutron stars is connected with its magnetic field \citep{Pons2007}.
Moreover, kinetic age determinations also show magnetic field decay \citep{Harrison1993}. Last studies find  that ages determined
by expansion of the parent supernova nebula are often smaller than those determined by the spin-down of the associated pulsar
\citep{ZhangXie2011}. Also pulsars are ofter closer to the galactic plane than it can be expected \citep{GuseinovAnkay2004}.
Though the population synthesis still leave problem unsolved \citep{SBPopov10, fauchergiguere06BirthEvolutionIsolatedRadioPulsars, IgoshevKholtygin2011}. 

Meantime, there is an additional effect of the magnetic field decay which still has not been carefully studied. In fact, changes in
magnetic field affect the spin-down age of radio pulsar. Many authors define the spin-down age as:
\begin{equation}
\label{tau_def}
\tau = \frac{P}{2\dot P}
\end{equation}
This formula estimates the age of the pulsar based on assumption that one steadily increases its period P from 0 to current
value. Although this quantity is commonly used, we understand its properties poorly. Indeed, this estimation should be the same as the real age 
only for quite old pulsars (with ages more than $10^4$ years). Moreover it strictly depends on magnetic field decay and alignment. 
For instance, if B, magnetic field strength, decreases following a power low with exponent $\gamma$, the spin-down age increases as
$\tau = \gamma t$ (see the second Section for deductions). 

Nevertheless, B is exchanged to $\tau$ does not give us additional information. We still need to observe $\tau (t)$ for every particular pulsar to restore magnetic
field history. This difficulty can be avoided if we assume a uniform magnetic field decay law $B = B_0 f(t)$ where $B_0$
is the initial magnetic field and $f(t)$ - is universal time-dependent law. This assumption leads us to conclusion that
all pulsars can be sorted by $\tau$. It intends that the higher $\tau$ is, the higher real age is also. Therefore we can consider pulsar with particular
age $\tau$ as a stage in the life of every pulsar in the Galaxy.  

After all, it is appeared that $\tau$ is $B$-dependent and vice verse. So, does it intend that we lost only available age estimation?
Probably not.  Let us have a look to different process which like annual rings in cross-section cut of a trunk deliver us age. 
It is quantity of pulsars. This quantity is determined by a birthrate, conditions which leads to pulsars detection and life-span of them. 
The first of them is number of pulsars which on average are born during fixed period of time. It is quite constant value which 
hardly ever variate significantly due to fixed life-span of massive stars and absence of starbursts in our Galaxy. 
Two other values strongly depends on pulsars characteristics. However, on average for ensemble it is seen equilibrium. 
Then let us continue analogy with annual rings. Every one thousand years new pulsars are born as a new annual ring during summer. 
When we count quantity of pulsars with some particular spin-down age we notice lacks and excesses which mark changes in
magnetic field.


Consequently, there are two independent age estimate $\tau$ and $t_{\mathrm{aver}}$ (deduced from birthrate of pulsars).
The spin-down age $\tau$ is affected by changes
in $B$. Nevertheless, many readers can suggest a few reasons why $\tau$ is bigger than $t$ even without magnetic field decay. For instance,
we observe smaller quantity of pulsars with bigger $\tau$ because they are weaker. Although this effect is significant, it can be avoided if we take
into consideration a part of our Galaxy. Also effect of alignment can increase the spin-down age. Both of these two effects
will be discussed in detail.

The article is structured as follows. The second Section is devoted to clarification of some important features of $\tau$. 
In the third Section we discuss uniform law of magnetic field decay and consequences of this assumption. The fourth Section
is devoted to a derivation of kinetic equation which allows us to restore real ages of pulsars. In the fifth and seventh Sections 
we present discussion of key selection effects. The method and results are presented in the sixth Section. 

\section{Mathematical properties of $\tau$}
\subsection{Connection with magnetic field}
Let us write the standard expression for magnetic field if a pulsar is braked due to electromagnetic radiation:
\begin{equation}
\label{field}
\alpha B^2 = P \dot P
\end{equation}
Where $\alpha$ is a constant for the given pulsar at given moment which depends on mass, momentum of inertia and angle $\xi$ between magnetic
 pole and the axis of rotation. 
In early work by \citet{jpostriker69NaturePulsarsITheory} the authors considered pulsar braking as a consequence of dipole radiation emission
in vacuum. However, \citet{gurevich93PhysicsPulsarMagnetosphere} showed that the pulsar has to posses an extended magnetosphere filled with plasma.
Therefore mechanism  \citep{jpostriker69NaturePulsarsITheory} does not work in itself. In fact, \citet{gurevich93PhysicsPulsarMagnetosphere} deduced
similar formula (\ref{field}) as \citet{jpostriker69NaturePulsarsITheory} did, except the angular dependence. The former 
believes that $\alpha \sim \cos^2(\xi)$ when the latter supposes that $\alpha \sim \sin^2(\xi)$.

On the other hand astronomers hardly ever measure both B and $\xi$ in their observation. The angle $\xi$ is measured by behavior
of polarization angle during a pulse \citep{Manchester1971, Lyne1971, McCulloch, Morris1981}.  The direct magnetic
field measurements are possible by detection of synchrotron absorption features in high energy part of spectra \citep{Blandford1976}.
However, this measurements tend to differ from quantity (\ref{field}) at some constant 
factor. It is might connected with different heights where high energy and radio emission photons are born. The height is smaller,
 magnetic field is stronger. In the paper \citep{TaurisManchester} it is shown that there is a relation although it does not 
behave like $\sin^2\xi$ or $\cos^2 \xi$, it is much more weaker. Therefore the conclusion about angular dependence can not be deduced from
observations right now.


Meantime, in observations we see that all isolated pulsars are braking. Therefore we suppose angular dependence 
in the following form:
\begin{equation}
\label{ang_dep}
(a\cos^2\xi+b\sin^2\xi)\alpha ' B^2 = (a + (b-a)\sin^2\xi)\alpha ' B^2=P\dot P
\end{equation}
This expression means constant braking rate with weak angular dependence ($|b-a| < a$). For sake of simplicity elsewhere in this article
except the seventh section we use $a=b$. 

Meanwhile no angular dependence involves no alignment. The alignment is a phenomenon when $\xi$ decreases with time.
Therefore taking into consideration expression (\ref{field}) in form (\ref{ang_dep}) we assume that there is no
alignment or its effect is negligible on interval of time shorter than $10^6$ years.

The expression (\ref{field}) is not so easy-to-use as integrated expression for purposes of our article. We can treat (\ref{field})
as a differential equation, and its solution is:
\begin{equation}
\label{int_form}
\frac{P^2(t)}{2} = \int _0 ^t \alpha (\tau ')B^2(\tau ') d\tau ' + C
\end{equation}
Where $C$ can be found from boundary condition $P(0)=P_0$ where $P_0$ is the initial period of the pulsar. Let us combine
(\ref{tau_def}) and (\ref{int_form}):
\begin{equation}
\label{int_form_final}
\tau(t)=\frac{\int _0 ^t \alpha (\tau ')B^2(\tau ') d\tau ' + 0.5 P_0 ^2}{\alpha(t)B^2(t)}
\end{equation}
This is an integral expression for the spin-down age which helps us to study its properties.

\subsection{The simplest magnetic decay models}
Currently astronomers tend to rely on a power-law and exponential magnetic field decay models. The first describes behavior of the field in a form: 
 
\begin{equation}
\label{B_power_low}
B=B_0 \left(\frac{t}{t_0}\right)^{\gamma}
\end{equation}
Using (\ref{int_form_final}) we find the spin-down age behavior:
\begin{equation}
\label{res_power_low}
\tau(t)=\frac{\int _0 ^t \tau '^{2\gamma} d\tau '}{t^{2\gamma}} + \frac{P_0 ^2}{2\alpha B(t)} = \frac{1}{2\gamma+1} t + \frac{P^2_0}{2\alpha B^2(t)}
\end{equation}
Clearly seen that the spin-down age in this model is proportional to real age and the rate is constant. 
More often astronomers use the exponential magnetic field decay law in pulsar population e.g. \citet{SBPopov10}. Let us suppose that the field decays as following:
\begin{equation}
\label{B_expon}
B=B_0 e^{-t/\tau}
\end{equation}
Then we apply relation (\ref{int_form_final}):
\begin{equation}
\label{res_expon}
\tau(t)=\frac{\int _0 ^t e^{-2\tau'/\tau} d\tau '}{e^{-2t/\tau}} + \frac{P_0 ^2}{2\alpha B^2(t)} = \frac{\tau}{2} - \frac{\tau}{2}e^{2t/\tau} + \frac{P^2_0}{2\alpha B^2(t)}
\end{equation}
Here the spin-down age is growing exponentially. These two the simplest examples hint that the spin-down age
is highly sensitive to behavior of $B$ and is weakly affected by the initial period and initial magnetic field.
It explains why we choose this quantity in our study.


\subsection{Averaging}
About 1700 isolated pulsars are known in our Galaxy. Each of them has specific parameters set. This set contains
magnetic field strength, period, period derivative, radio-luminosity, rotation axis orientation etc.
Some of the characteristics are connected with each other. For instance, formula (\ref{field}) connects period,
derivative of period and magnetic field. Using this formula we can deduce any of this characteristics by others.

There are characteristics which are not connected with each other. They are independent. For instance, nowadays
we can not suggest how initial magnetic field of a neutron star and initial period are related. Therefore let us call two
values independent further if we can not suggest any physical law which associates them and we do not see any correlation
in observational data.

The ensemble of pulsars contains about 1700 discrete objects. Every object there has the whole set of observed 
characteristic. Thereby, we observe the discrete set of characteristics.
Meantime,  there is no physical argument in favor that it is not possible for an additional pulsar
to have some intermediate characteristics. It means that observed ensemble is a realization of some distribution.
Therefore it is quite natural to replace one by continuous distribution and work further with
it instead of particular pulsars.

Let us suppose that $\zeta$ is a property of a pulsar such as a period, magnetic field or something like this. Then we introduce
the distribution function $\omega(\zeta)$. Let us choose it as a normalized one:
\begin{equation}
\label{norm}
\int_a^b \omega(\zeta)d\zeta = 1
\end{equation}
When we average some quantity on this distribution we get an expectation value of this quantity. Due to presence of variety
in observation data for every parameters, we  use expectation value in our analysis.

\subsubsection{Averaging on initial magnetic fields distribution}
As it is mentioned above the initial magnetic fields distribution and initial periods distribution are independent.
Therefore we can average on these characteristics independently. Let us consider  $B_0$ as $\zeta$. We also
introduce here the following designation: 
\begin{equation}
\label{aver_B}
\overline{\tau(t)}|_{B_0} = \int _{B_1} ^{B_2} \tau(t, B'_0) \omega(B_0')d B_0'
\end{equation}
Then the expression (\ref{int_form_final}) can be rewritten:
\begin{equation}
\label{aver_B_1}
\overline{\tau(t)}|_{B_0} = \overline {\left.\frac{\int _0 ^t \alpha (\tau ')B^2(\tau ') d\tau ' + 0.5 P_0 ^2}{\alpha(t)B^2(t)}\right|}_{B_0}
\end{equation}
The initial magnetic field might affects $\alpha$ only during first about $10^3$ years when it significantly deforms neutron
stars if the field is strong enough \citep{ThompsonDincan2000, Ghosh2011, Ghosh2009, Haskell2008, Colaiude2008, Wasserman2003, jpostriker69NaturePulsarsITheory}.
On the contrary the aim of our investigation are 
middle-aged pulsars. Therefore we can carry $\alpha$ out of the integral. Moreover as any physical quantity $B(t)$ and
$\omega (B_0)$ are continuous, so it is possible to exchange integrals:
\begin{equation}
\label{aver_B_2}
\overline{\tau(t)}|_{B_0} = \int _0 ^t \int_{B_1}^{B_2} \frac{B^2(\tau ', B_0)}{B^2(t, B_0)} \omega (B_0) dB_0 d \tau ' + \frac{P^2_0}{2\alpha \overline {B^2}}
\end{equation}
Here we take assumption about behavior of magnetic field. We suppose that $B(t) = B_0 f(t)$, and $f(t)$ is a monotony function. Thereby:
\begin{equation}
\label{aver_B_3}
\overline{\tau(t)}|_{B_0} = \int _0 ^t \int_{B_1}^{B_2} \frac{f^2(\tau ')}{f^2(t)}\omega (B_0) dB_0 d \tau ' + \frac{P^2_0}{2\alpha \overline {B_0^2}f^2(t)}
\end{equation}
Because $f(t)$ does not depend on the initial magnetic field and using relation (\ref{norm}) we take:
\begin{equation}
\label{aver_B_4}
\overline{\tau(t)}|_{B_0} = \frac{\int_0^t f^2(\tau ')d\tau '}{f^2(t)}  + \frac{P^2_0}{2\alpha \overline {B_0^2} f^2(t)}
\end{equation}
It is again the spin-down age with a small disturbance. This disturbance will be estimated after the averaging on the initial periods 
has been done. 

\subsubsection{Averaging on initial periods distribution}
Let us do here the averaging on initial periods distribution  similarly as it is done on initial magnetic fields in the previous section. 
Firstly, consider $\zeta$ as $P_0$. We designate the averaging by:
\begin{equation}
\label{aver_P}
\overline{\tau(t)}|_{P_0} = \int _{P_1} ^{P_2} \tau(t, P'_0) \omega(P_0')d P_0'
\end{equation}
Using the relation (\ref{int_form_final}) we can rewrite:
\begin{equation}
\label{aver_P_1}
\overline{\tau(t)}|_{P_0} = \overline {\left.\frac{\int _0 ^t \alpha (\tau ')B^2(\tau ') d\tau ' + 0.5 P_0 ^2}{\alpha(t)B^2(t)}\right|}_{P_0}
\end{equation}
The initial period might affects on $\alpha$ only during first about $10^3$ years when it significantly deforms neutron crust
if it is very short \citep{jpostriker69NaturePulsarsITheory, Cutler2003}, after this time gravitational radiation brakes neutron stars enough that
changes in $\alpha$ can be neglected. As it is mentioned above the aim of our investigation are middle-aged pulsars. Therefore we can
carry $\alpha$ out of integral due to $\alpha$ is constant on our time-scale. Moreover, as any physical quantity, $P(t)$ and $\omega (P_0)$
are continuous, so it is possible to exchange integrals:
\begin{equation}
\label{aver_P_2}
\overline{\tau(t)}|_{P_0} = \int _0 ^t \int_{P_1}^{P_2} \frac{B^2(\tau ')}{B^2(t)} \omega (P_0) dP_0 d \tau ' + \frac{\overline{P^2_0}}{2\alpha B^2}
\end{equation}
As it is mentioned above initial magnetic field and initial period are not connected. Therefore we can carry out $\int _0^t \alpha B^2(\tau ')$ from
$\int _{P_1}^{P_2}  dP_0$:
\begin{equation}
\label{aver_P_3}
\overline{\tau(t)}|_{P_0} = \frac{\int_0^t f^2(\tau ')d\tau '}{f^2(t)}  + \frac{\overline {P^2_0}}{2\alpha B_0^2 f^2(t)}
\end{equation}
And again we get similar the spin-down age with small disturbance.

\subsubsection{Averaging on both distributions}
Now, when we already know expressions (\ref{aver_B_4}) and (\ref{aver_P_3}), we average on both characteristics
simultaneously:
\begin{equation}
\label{both} 
\overline{\overline{\tau(t)}|_{P_0}}|_{B_0} = \overline{\overline{\tau(t)}|_{B_0}}|_{P_0} := \overline{\tau (t)}|_{P_0,B_0}
\end{equation}
The result is:
\begin{equation}
\label{both_1}
\overline{\tau(t)}|_{P_0,B_0} = \frac{\int_0^t f^2(\tau ')d\tau '}{f^2(t)}  + \frac{\overline {P^2_0}}{2\alpha \overline {B_0^2} f^2(t)}
\end{equation}
We should carry $f(t)$ out of parenthesizes because it affects similar on the first and the second summand. Then we take typical values
for members in numerator and denominator from \citep{fauchergiguere06BirthEvolutionIsolatedRadioPulsars} as $P_0=0.1$ sec, $B=4\cdot 10^{12}$ G and 
$\alpha = 3.2\cdot 10^{-19}$.  Therefore this disturbance member can be estimated as:
\begin{equation}
\label{est_dist}
\frac{\overline {P^2_0}}{\alpha \overline {B_0^2}}=\frac{10^{-2}}{10^{-39}\cdot 4\cdot 10^{25}} \approx 2.5\cdot 10^{11} \; \mathrm{sec} \approx 8\cdot 10^3 \; \mathrm{years}
\end{equation}
It means that $\overline \tau |_{B_0, P_0}\sim \tau$ without any disturbance members for quite old pulsars. Consequently
our method can be applied for pulsars with the spin-down age which is larger that $4\cdot 10^4$ years.
This value defines boundary where our method can be applied.

\section{The uniform magnetic field decay law}
In order to clarify how physical the assumption about the uniform law is we need to consider what properties of a neutron star can affect the magnetic field decay.
These properties are rotation, crust conductivity, accretion, temperature and properties of
magnetic field itself such as configuration and strength. Let us discuss these factors. 

If we exclude objects which accrete matter, we consider only pulsars with similar properties. 
Indeed, majority of neutron stars are born with short periods with small variance
\citep{fauchergiguere06BirthEvolutionIsolatedRadioPulsars}.
Neutron stars are born in supernovae explosions, and consequently should have similar characteristics in beginning
because of similar conditions of supernova explosion. We exclude accretion since 
we take into consideration only isolated pulsars with quite long periods.
As for magnetic field configuration, pulsars tend to brake such way that dipole magnetic field configuration
is looked as the most probable. Nethertheless, pulsars with the strongest magnetic fields brake in a different way. They
emit radiation due to destruction of magnetic field. These pulsars are magnetars. However, we exclude 
them from our selection either. Therefore this assumption about the uniform magnetic field decay law seems
quite natural. 

Moreover, this assumption also gives us time order. Indeed, if we have two pulsars one with the initial magnetic
field $B_1$ and another one with $B_2$, then for every time $t_1>t_2$ $\tau _1> \tau _2$ and vice verse.
Indeed:
\begin{equation}
\label{abs_tau}
\tau_1 = \frac{\int_0^{t_1}\alpha B_0^2 f^2(t')dt'}{\alpha B_0^2 f(t_1)} = \frac{\int_0^{t_2} f^2(t')dt'}{f(t_1)} + \frac{\int_{t_2}^{t_1} f^2(t')dt'}{ f(t_1)}
\end{equation}
$f(t_1)<f(t_2)$ ($f(t)$ is monotonous function) it means that the first summand is bigger or the same as $\tau_2$.
Simultaneously the second summand is positive, and consequently $\tau_1>\tau_2$

When we study the ensemble of real pulsars we should remember that there is a condition when a pulsar turns off. 
This condition is determined by combination of $B$ and $P$. It means that pulsars with lower fields turn off
earlier. Due to this fact, our selection is not complete for larger ages.  

Currently the death-line for isolated radio-pulsars is determined by following condition \citep{Bhattacharya1992}:
\begin{equation}
\label{turn_off}
\frac{B}{P^2}<c
\end{equation}
Here $c$ is a constant. On the other hand we can rewrite this expression via $B$ and $\tau$:
\begin{equation}
\label{turn_off_tau}
\tau(t)>\frac{1}{2\alpha B_0 c f(t)}
\end{equation}
Then we estimate $B_0=10^{13}$G, $c=1.7\cdot 10^{11}$G s${}^{-2}$, $f(t)=0.1$ and get $\tau_{\mathrm{turn-off}}=10^8$ years. However,
near $10^6-10^7$ years effects connected with alignment begin influencing, see \citet{Young2010} and references therein.

\section{Kinetic equation and statistical age estimation}
For the whole pulsar ensemble we can write a kinetic equation. 
Let us consider small bits at the spin-down age and real age space. In this two dimensional space we consider
real ages as time-coordinate and the spin-down ages as space-coordinate. This approach differs from that suggested by
\citet{BeskinGurevich1986, Phinney1981, Vivekanand1981, Deshpande1995} 
where authors considered distribution function out of $(P,B,\xi)$. In our work we suppose that there is no apparent 
angular dependence, the magnetic field $B$ changes due to uniform law $B=B_0 f(t)$. 
Therefore $\tau$ is a function only of $t$ for quite old ages (see the second Section for deductions).
In these terms $n(\tau, t)$ is pulsars distribution function. It denote quantity of pulsars with the spin-down from
$\tau$ to $\tau + d\tau$ and real age from $t$ to $t+dt$. We can write now the continuity equation for the whole pulsar ensemble:
\begin{equation}
\label{cont_equat}
\frac{\partial n}{\partial t} + div\left(n\frac{d\tau}{dt}\right)=U-V
\end{equation}
Here $U$ is a source of pulsars and $V$ is a summand which describes pulsars death. The simplest source of pulsars in
this coordinate space should be $n_{\mathrm{birthrate}} \delta(0)$ which sets the spin-down age for all new-born pulsars as zero. 
In reality, majority of pulsars are born with the spin-down age near $10^3$ years. And the total birth rate is determined as:
\begin{equation}
\label{R}
R = \int _{B_{\mathrm{min}}}^{B_{\mathrm{max}}} \int _{P_{\mathrm{min}}}^{P_{\mathrm{max}}} \rho (B_0) \rho (P_0) dB_0 dP_0
\end{equation}
Here $\rho(B)$ determines the initial distribution of pulsars by their magnetic fields, and $\rho(P_0)$ by their
initial periods.
We can define parametrization $B_0=B_0(\lambda, \tau)$ and $P_0=P_0(\lambda, \tau)$ which determines this way
that $\forall \lambda$ $P_0^2/(2\alpha B_0^2) = \tau$. Then we integrate $\rho(B_0)\rho(P_0)$ by this parameter $\lambda$:
\begin{equation}
\label{R_tau}
\rho(\tau) = \int _{\lambda_{\mathrm{min}}}^{\lambda_{\mathrm{max}}}  \rho (B_0(\lambda)) \rho (P_0(\lambda)) d\lambda
\end{equation}

\begin{equation}
\label{R_ch_var}
R = \int _{0}^{\tau_{\mathrm{max}}} \rho (\tau) d\tau
\end{equation}

It is important to mention that there is the highest possible value of $\tau$ at the birth. And pulsars with $\tau>4\cdot 10^4$ years
at the birth are rare. Therefore we can truncate this distribution function at value $4\cdot 10^4$ years.  Because we
take into consideration pulsars with the spin-down ages smaller than $2\cdot 10^6$ years we can neglect $V$.
We also suppose that the pulsars ensemble is stationary. 
Following these assumption we can rewrite (\ref{cont_equat}):
\begin{equation}
\label{cont_equat_1}
\frac{\partial n}{\partial \tau} \frac{d\tau}{dt} =  \rho (\tau) \frac{dB}{d\tau}  
\end{equation}
In order to find $n(\tau, t)$ in a certain position we should integrate this relationship:
\begin{equation}
\label{cont_equat_2}
\int _0 ^{\tau} \frac{\partial n}{\partial \tau} d\tau = \int _0 ^{\tau}  \rho(\tau) \frac{dt}{d\tau}d\tau
\end{equation}
The result is:
\begin{equation}
\label{cont_equat_3}
n (\tau, t) =  n_{\mathrm{birthrate}}\frac{dt}{d\tau}
\end{equation}
It is important to remind that $n$ is density, so the quantity of pulsars with the spin-down age smaller than $\tau$ is:
\begin{equation}
\label{def_n}
N (\tau, t) =  \int_0 ^{\tau} n(\tau, t) d\tau
\end{equation}
Let us put (\ref{def_n}) into (\ref{cont_equat_3}) and integrate the result:
\begin{equation}
\label{result}
t(\tau) = \frac{N(\tau)}{n_{\mathrm{birthrate}}}
\end{equation}

\section{Selection effects}
Results of the method, either it is right or not, might be spoiled due to significant observational selection. In fact,
we observe only a small part of all galactic pulsars. And we do not have any solid base to suppose that this small part correctly
reproduces characteristics of the whole galactic ensemble. Indeed, selection effects are able to hide some group of pulsars from the whole ensemble.
In this case selection effects change an average and a variance of the observed distribution. 

For purposes of our study the most important selection  is that which is caused by decrease of radio luminosity for old pulsars. Astrophysics suppose
that radio luminosity of pulsars is a constant part of their rotational energy. Due to this hypothesis old pulsars
with longer period should have smaller supply of kinetic energy, and consequently they should be weaker radio sources. 
We are able to observe weaker sources only closer to the Sun. Moreover, following article \citep{gurevich93PhysicsPulsarMagnetosphere}
pulsars with higher magnetic fields brakes quicker, and consequently have smaller luminosity. It might cause 
apparent magnetic field decay even if it does not have place. Our method should consider full selections
for all spin-down ages before pulsars shutdown. Therefore this effect demands careful consideration.

\begin{figure*}
\begin{minipage}{0.47\linewidth}
\center{\includegraphics[width=1\linewidth]{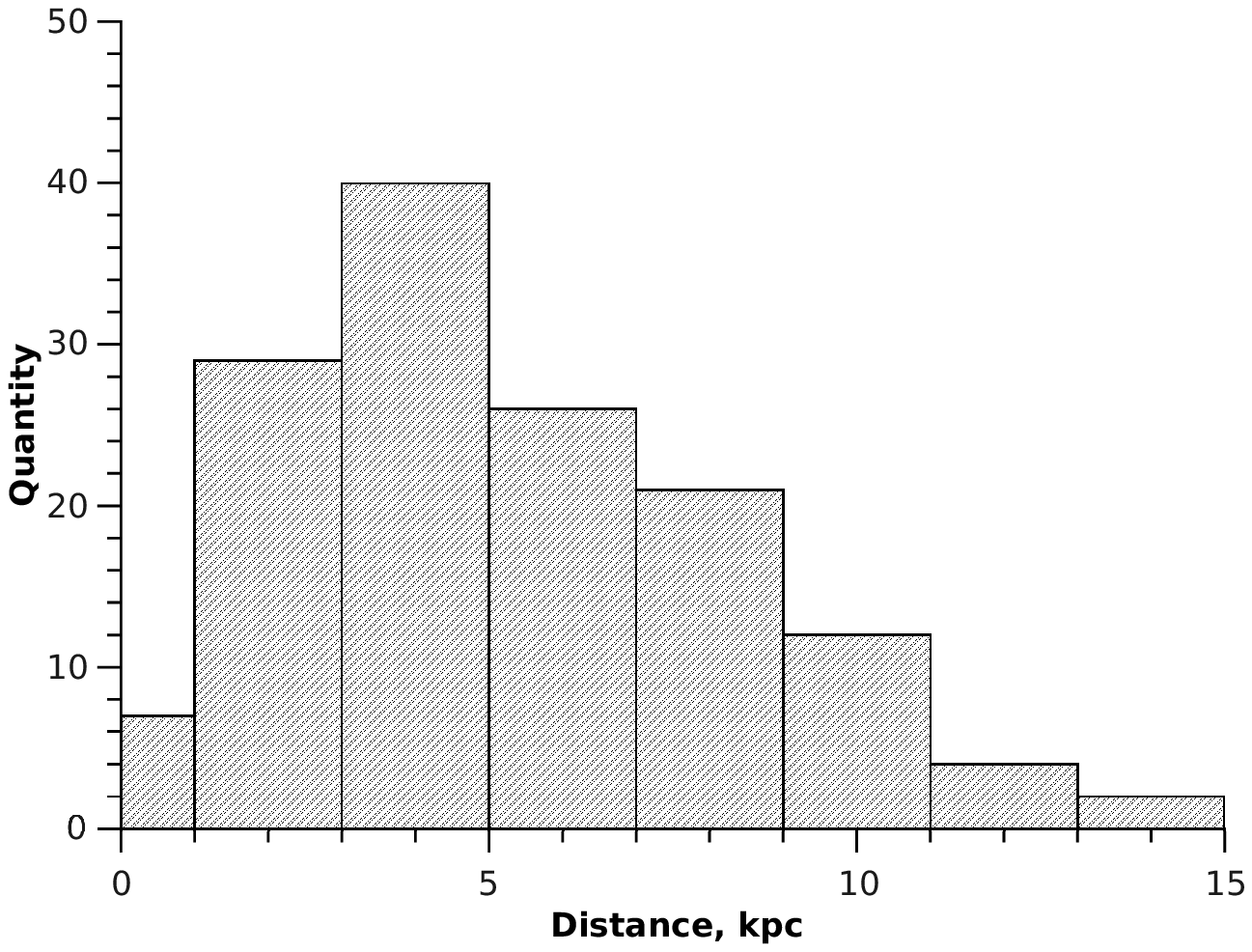}} a) \\
\end{minipage}
\hfill
\begin{minipage}{0.47\linewidth}
\center{\includegraphics[width=1\linewidth]{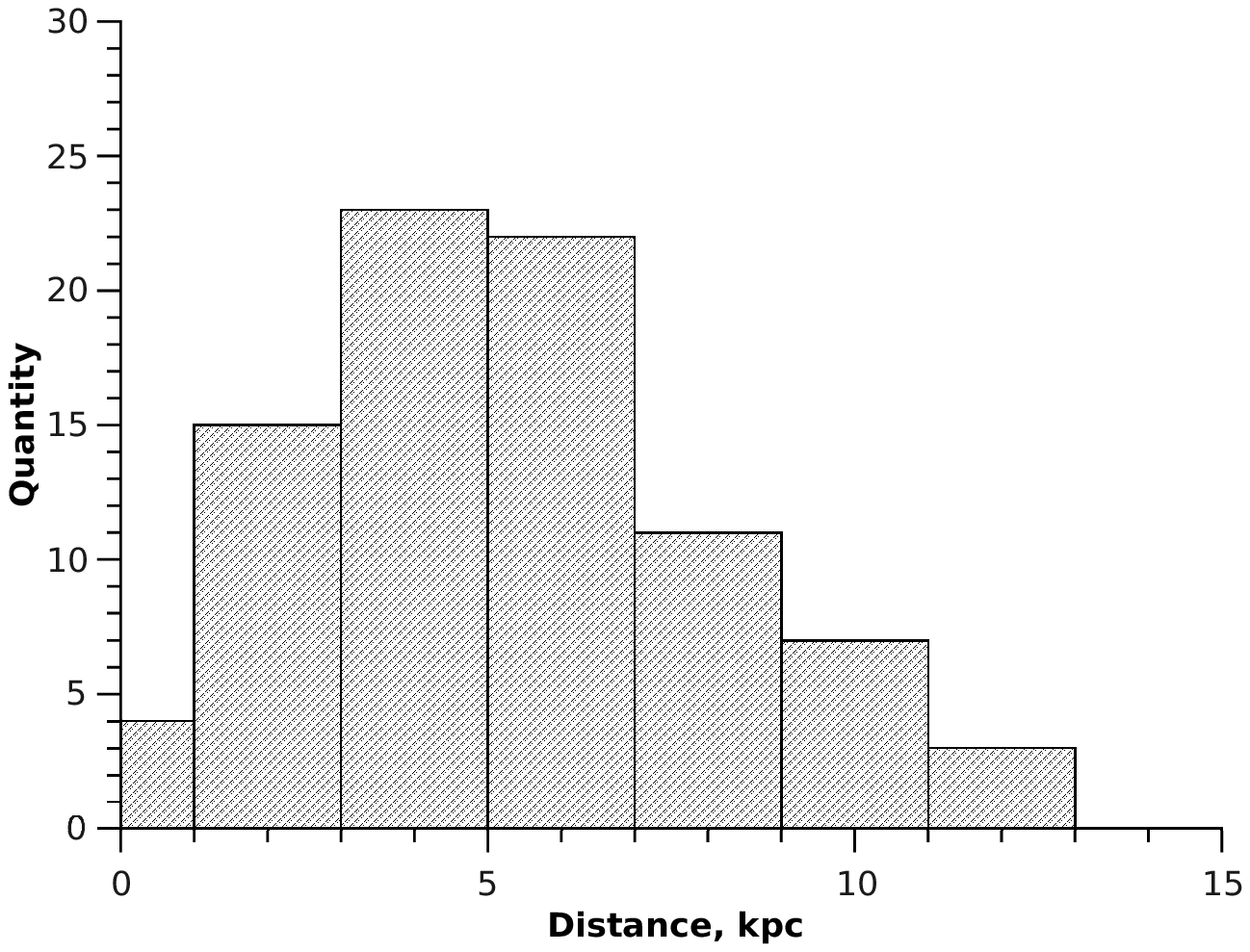}} b) \\
\end{minipage}
\vfill
\begin{minipage}{0.47\linewidth}
\center{\includegraphics[width=1\linewidth]{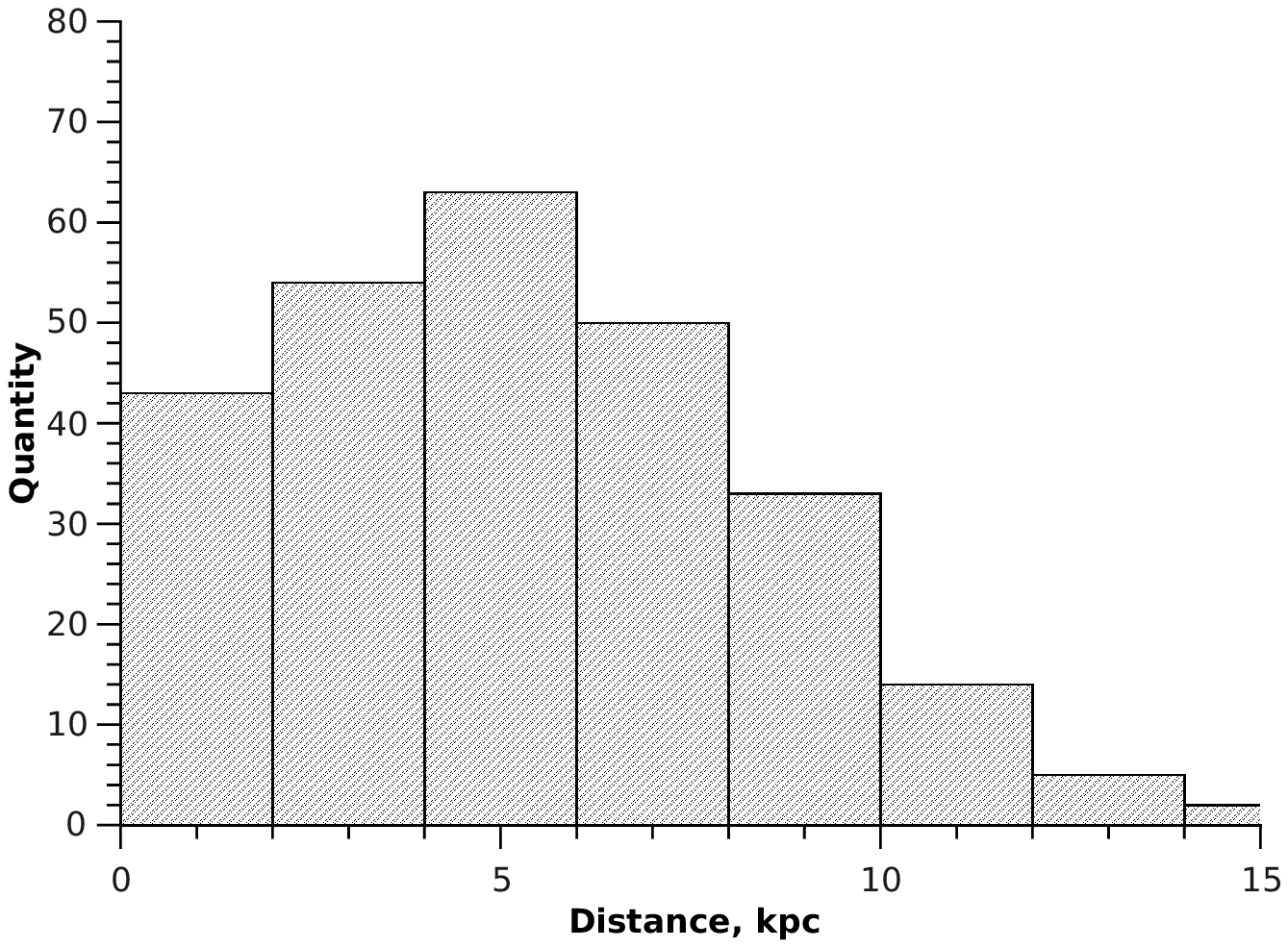}} c) \\
\end{minipage}
\hfill
\begin{minipage}{0.47\linewidth}
\center{\includegraphics[width=1\linewidth]{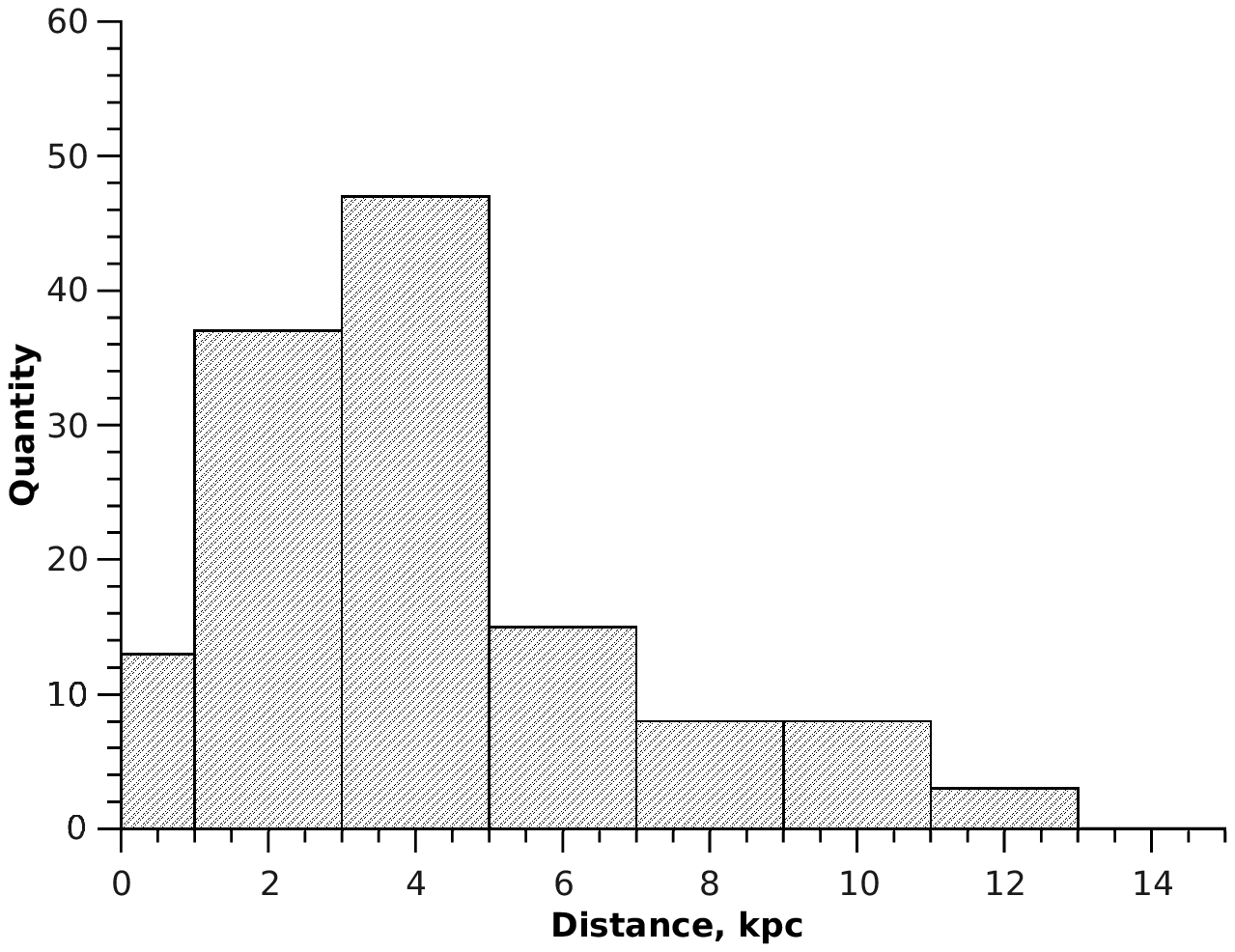}} d) \\
\end{minipage}
\caption{
Distributions of pulsars by distance for different intervals 
of the spin-down ages a) $10^3-2\cdot 10^5$ years, b)
$2\cdot 10^5-5\cdot 10^5$ years, c) $5\cdot 10^5-2\cdot 10^6$ years, d) 
$4\cdot 10^7-10^8$ years}
\label{obser_select}
\end{figure*}

On the other hand it is difficult to estimate this selection on the theoretical ground because of our poor understanding
of the radiation mechanism. Therefore we draw our attention to observational data. We choose isolated radio-pulsars
which are not millisecond ones from the ATNF pulsar catalogue \footnote{http://www.atnf.csiro.au/research/pulsar/psrcat/} \citet{manchester05ATNFPulsarCatalog}. 
We divide them into four groups: (the spin-down age is in years) $\tau\in [10^3-2\cdot 10^5]$,  $\tau\in [2\cdot 10^5-5\cdot 10^5]$, $\tau\in [5\cdot 10^5-2\cdot 10^6]$
and $\tau\in [4\cdot 10^7- 10^8]$ and plot them in Figure \ref{obser_select} due to their distance from the Sun. It is clearly seen
that shares first three of them are similar when we consider distances which smaller than 10 kpc and the last shows evidences of observational selection.
Consequently, this procedure lets us to avoid observation selection.


\section{Description of the method}
\begin{figure}
\center{\includegraphics[width=1.\linewidth]{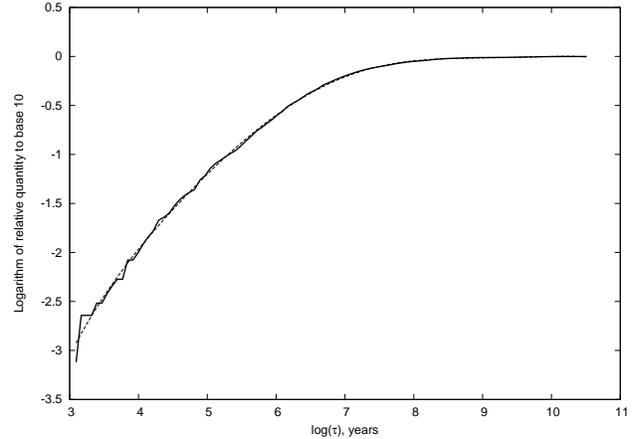}}
\caption{Cumulative pulsar distribution by the spin-down ages. Solid line - distribution of pulsars and dashed line is approximation.}
\label{distr_cum}
\end{figure}

The first step of our method is to plot $n(\tau)$ as a cumulative distribution. This distribution is presented
in the Figure~\ref{distr_cum}. When we plot this figure we extract from the ATNF pulsar database only
isolated non-millisecond pulsars  which lie closer than 10 kpc to the Sun. It allows us to avoid 
observational selection (see the fifth Section for details) and effects of accretion. Therefore both
our assumptions about the uniform magnetic field decay law and fullness of selection seem correct. Then
we exchange discrete distribution to continuous one by approximation of polynomial, and treat this 
distribution as $\overline {\tau}|_{P_0,B_0}=\tau$ the universal spin-down distribution (see the second
Section for proof). We choose polynomial of 6th degree to avoid too detailed, wave-like behavior near
$10^3-10^4$ years and on the other hand to carefully reproduce the form of distribution from
$10^4-2\cdot 10^6$ years. We fit distribution by last-square method and get following formula:
\begin{equation}
\label{app_polinom}
g(x) = w x^6 + a x^5 + b x^4 + c x^3 + d x^2 + f x + g
\end{equation}
With coefficients $a=0.00883$, $b=-0.14894$, $c=1.2942$, $d=-6.2086$, $f=16.54$, $g=-21.665$ and $w=-2.102\cdot 10^{-4}$

The second step of our method is conversion from $n(\tau)$ to $t(\tau)$. This can be done accordingly to
the fourth part of our article. This conversion needs some estimate of pulsar birth-rate. 
We do this estimate rely on consequences of our model field decay. We suppose that $f(t)$ is a monotony function.
Therefore for every moment of time $\tau \geq t$. We set $\tau=t$ at the moment $4\cdot 10^4$ years and calculate birthrate 
due to this assumption.
Then we relay on fact that it is possible to observe only $10\%$ of all
pulsars and comparing volume of 10 kpc Sun-centered circle to the whole volume of the Galaxy we recalculate birthrate from $1.1$
 visible pulsars per thousand of years to $2.9$ pulsars per century.
 It looks like a minimum
estimate because in reality it appears that $\tau(t) \gg t$. However, some astronomers believe that the magnetic field decay
occurs on time-scale much more than $10^4$ years. It means that when age of pulsar is about $10^4$ years his real and the spin-down ages are similar. 
Therefore this estimate should give the real birthrate of pulsars in the Galaxy. Following this method we get $n_{\mathrm{birthrate}}=2.9$ pulsars
per century which is slightly greater than \citet{fauchergiguere06BirthEvolutionIsolatedRadioPulsars} got. 
This birth-rate is
two times greater than 1.4 pulsars per century from \citet{Vranesevic2004} estimate. Nethertheless our estimation also
agrees with the rate ofsupernova explosions \citep{Keane2008}. 

\begin{figure}
\center{\includegraphics[width=1.\linewidth]{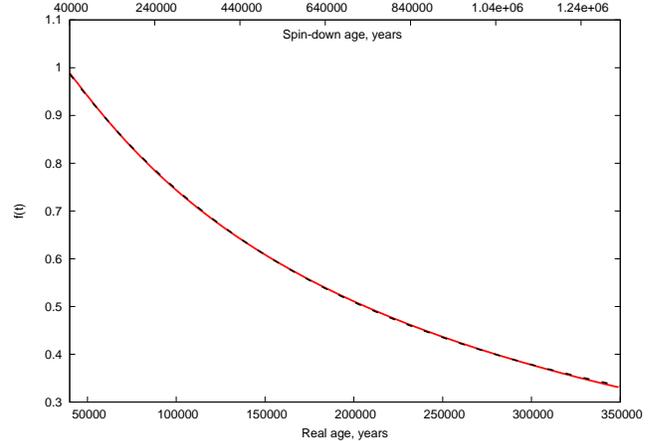}}
\caption{The magnetic field decay law $f(t)$ (solid line) for middle-aged radio pulsar with approximation (dashed line). }
\label{res_fin}
\end{figure}

Let us write here the equation (\ref{both_1}):
\begin{equation}
\label{diff_f_1}
\tau (t) = \frac{\int_0 ^t f^2(t')dt'}{f^2(t)}
\end{equation}
Let us carry $f(t)$ from denominator to left part and differentiate by $t$:
\begin{equation}
\label{diff_f_2}
2\dot f(t) \tau (t) + f(t) \cdot (\dot\tau (t)-1) =0
\end{equation}
Here it is possible to separate variables:
\begin{equation}
\label{diff_f_3}
\frac{\dot f(t)}{f(t)}=-\frac{\dot \tau (t)}{2\tau (t)} + \frac{1}{2\tau(t)}
\end{equation}
It can be integrated:
\begin{equation}
\label{diff_f_4}
f(t)=\frac{\exp(\int_0^t \frac{dt'}{2\tau (t')})}{\sqrt{\tau(t)}}
\end{equation}
Accordingly to this relationship we numerically calculated values of $f(t)$ for every moment of time. The 
result is presented in Figure \ref{res_fin}. It is possible to approximate this curve by following expression:
\begin{equation}
\label{appr_res}
s(t) = \left(\left(a\frac{t}{t_0}\right)^{\gamma}+c\right)^{-1}
\end{equation}
Here $\gamma=1.17$, $a=0.034$, $c=0.84$, $t_0=10000$ years.

\section{Discussion of alignment}
Alignment is a phenomenon when the angle between the magnetic field pole and the rotational axis decreases with time.
It is extremely important for our work because strong alignment affects on the spin-down age. It even can
ruins our conclusion about absoluteness of $\tau$. It happens because pulsars with high $\tau$ might be 
young pulsars with $\xi \to 90^{\circ}$. Therefore we may mistakenly mix young pulsars with older ones.
Then we can not say that the spin-down ages distribution some way reflects magnetic field decay

In order to protect our method from the alignment supporters criticism we need to consider alignment as a part of modern
braking models. One of them is model of magneto-dipole radiation \citep{jpostriker69NaturePulsarsITheory}. In this model: 
\begin{equation}
\label{dipole_P}
P\dot P = \alpha B^2(t)\sin^2\chi
\end{equation}
There is an invariant which remains constant during angle evolution \citep{EliseevaPopov2006}:
\begin{equation}
\label{dipole_I}
\frac{\cos\chi}{P}=I_{dp}
\end{equation}
Using commonly known relationship $\sin^2\xi+\cos^2\xi=1$ we can rewrite (\ref{dipole_P}), (\ref{dipole_I}) into differential equation:
\begin{equation}
\label{full_diff_dipole}
P(t)\dot P(t) + I_{dp}^2P(t)^2\alpha B^2(t) = \alpha B^2(t)
\end{equation}

\begin{figure}
\center{\includegraphics[width=1.\linewidth]{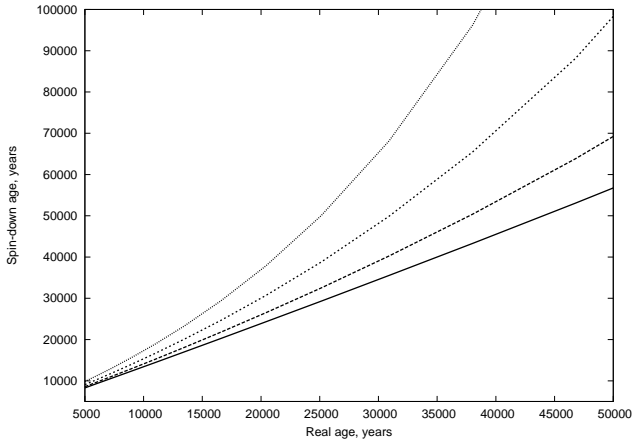}}
\caption{The spin-down pulsars evolution plotted by their real age. Different lines refer to different initial angles. Solid line - the initial angle is $85^{\circ}$, dashed - $80^{\circ}$,
 big dots - $75^{\circ}$, small dots - $70^{\circ}$ }
\label{angle_dep}
\end{figure}

The numerical solution of this equation lets us understand the value of alignment. We calculated the spin-down ages due 
to equation (\ref{full_diff_dipole}) with typical initial magnetic field ($B_0 = 5\cdot 10^{12}$ G) and initial period
($P_0=0.1$ sec). As we can see from the Figure \ref{angle_dep} only pulsars with the initial angles greater than $75^{\circ}$ 
at $40000$ years have the spin-down age near this value. On the other hand, this work by \citet{jpostriker69NaturePulsarsITheory} supposes homogenous
angle distribution. Therefore only $(90-75)/90=16.7$ per cent of all pulsars appears in the area where we measure the birth-rate. 
However, our birthrate estimate is in good agreement with other works. If  \citet{jpostriker69NaturePulsarsITheory} had been right, we would have taken
six times smaller birthrate estimate. Therefore we believe that there is no alignment or its effect is negligible. 

\section {Conclusion}
Based on the assumption about the uniformity of magnetic field decay law we suggest a new method which allows us to
restore the behavior of magnetic fields on the surface of middle-age radio pulsars. We prove that this method
can be applied to analysis of statistical properties of $\tau$. We restore the magnetic field decay law
and find that it can be fitted by modified power-law (\ref{appr_res}).
The birthrate of pulsars in the Galaxy is estimated based on the assumption that $\tau \approx t$ at 40000 years.
This estimate is in good agreement with other ones. Moreover, $\tau (t)$ evolution is calculated taking 
into account alignment. It is found that if alignment had taken place, we would have taken at once three times
smaller birthrate by our method than it is. 

\section{Acknowledgments}
We thank the Australia Telescope National Facility for making their pulsar data base available.
We thank Saint-Petersburg University grants 6.38.73.2011 and 6.46.627.2012. We would like to express our gratitude  
to S.B.Popov for careful reading this manuscript and many fruitful remarks. We also thank A.F. Kholtygin, V.A. Urpin and K.A.Postnov for useful discussions.

\bibliography{myref}
\bibliographystyle{mn2e}

\end{document}